\documentclass[11pt]{article}
\usepackage{setspace}
\usepackage{caption}
\usepackage{amsfonts}
\usepackage{graphicx}
\usepackage{amsmath}
\usepackage{float}
\usepackage{comment}
\newcommand {\be}{\begin{equation}}
 \newcommand {\ee}{\end{equation}}
 \newcommand {\bea}{\begin{array}}
 
 \newcommand {\eea}{\end{array}}

\textwidth=6.5in \hoffset=-.75in \voffset=-1in \numberwithin{equation}{section}
\numberwithin{figure}{section}

\begin{document}

\begin{titlepage}
\vspace{1cm} 
\begin{center}
{\Large \bf {Magnetized Kerr/CFT Correspondence}}\\
\end{center}
\vspace{2cm}
\begin{center}
\renewcommand{\thefootnote}{\fnsymbol{footnote}}
Haryanto M. Siahaan{\footnote{haryanto.siahaan@unpar.ac.id}}
\\
Center for Theoretical Physics,\\Physics Department, Parahyangan Catholic University,\\
Jalan Ciumbuleuit 94, Bandung 40141, Indonesia
\renewcommand{\thefootnote}{\arabic{footnote}}
\end{center}

\begin{abstract}
We extend the conjectured Kerr/CFT correspondence to the case of extremal Kerr black holes immersed by an magnetic field, namely the extremal Melvin-Kerr black holes. We compute the central charge which appears in the associated Virasoro algebra generated by a class of diffeomorphisms that satisfies a set of boundary conditions in the near horizon of an extremal Melvin-Kerr black hole. Our results support the Kerr/CFT conjecture, where the macroscopic Bekenstein-Hawking entropy for an extremal Melvin-Kerr black hole matches the result obtained from a dual 2D CFT microscopic computation using Cardy formula. Interestingly, the dual CFT description could be non-unitary, due to the possibility of negative central charge.
\end{abstract}
\end{titlepage}\onecolumn 
\bigskip 

\section{Introduction}
\label{sec:intro}

The conjectured Kerr/CFT correspondence \cite{Guica:2008mu}
states that the quantum theory of gravity in the near horizon of extreme Kerr black holes, which subjects to several specific boundary conditions,
is holographic dual to a 2D (two dimensional)
chiral CFT (conformal field theory). In this conjecture, Cardy formula in the dual 2D CFT can reproduce the macroscopic Bekenstein-Hawking entropy of extremal Kerr black holes. To obtain the central charge which corresponds to the near horizon geometry of an extremal Kerr black hole, Guica, et al. \cite{Guica:2008mu} adopt the method by Brown and Henneaux in \cite{Brown:1986nw} to compute the central charge associated to an $AdS_3$ spacetime. Getting the corresponding central charge which is related to the near horizon geometry under consideration is an important step in Kerr/CFT correspondence prescription \cite{Guica:2008mu}, since this central charge is needed in the dual calculation of entropy using Cardy formula. Subsequently, the Kerr/CFT correspondence \cite{Guica:2008mu} was extended to many cases of extremal rotating black holes \cite{Kerr-CFT-rot}, even to the static charged ones \cite{RN-CFT}.
  
Later on, Castro, et al. \cite{castro} broadened the discussion of Kerr/CFT correspondence to the case of non-extremal Kerr black holes. Instead of looking for the warped and twisted product of $AdS_2\times S^2$ structure of the near horizon non-extremal Kerr black holes, the authors of \cite{castro} show the hidden conformal structure in the wave equation of scalar fields in the near region of Kerr black holes. Solving the mapping between Boyer-Lindquist coordinates $(t,r,\phi)$ and the conformal ones $(\omega^+ , \omega^-, y)$, the temperature in left and right movers CFT can be found. Assuming that the central charge computed in the extremal case can still be used in the non-extremal situation, again the Cardy formula in 2D CFT can reproduce the macroscopic Bekenstein-Hawking entropy for non-extremal Kerr black holes. This work was also extended afterwards into many cases of non-extremal rotating black holes, for example \cite{Hidden-Rot}, and to static charged ones as well \cite{Hidden-RN}.

Due to such an enormous magnetic fields in the nucleus of galaxies \cite{Eatough:2013nva}, the interaction between magnetic fields and black holes plays an important role in explaining some observed physical aspects of supermassive black holes in the center of galaxies. Furthermore, because of most supermassive black holes must be rapidly spinning \cite{Elvis:2001bn}, including the ones in the center of galaxies, then it would be interesting to see how the Kerr/CFT correspondence works in the case of black holes immersed in magnetic fields or embedded in a magnetic universe. Quite recently, some interesting discussions on the magnetized black holes have appeared \cite{Booth:2015,Gibbons:2013yq,Gibbons:2013dna}, where in particular the authors of \cite{Konoplya:2008hj,Brito:2014nja} discussed the instability of scalar fields in the background of black holes immersed in magnetic fields. These works motivate us to study the Rotating Black Holes/CFT correspondence analysis, at the moment is focused on the extremal case, to the case of black holes immersed by an external magnetic field.  As a matter of fact, the author of \cite{Astorino:2015lca} shows how the  Reissner-Nordstrom (RN)/CFT can be extended to the case of magnetized RN/CFT correspondence, i.e. performing holographic calculation in 2D CFT using Cardy formula to reproduce the Bekenstein-Hawking entropy of extremal magnetized-RN black holes\footnote{We learn that the author of \cite{Astorino:2015lca} is also preparing a study that potentially overlaps with the analysis presented in this manuscript.}. 

The obtained results in this paper are not trivial. In addition that Kerr/CFT correspondence still holds in the case of a magnetized Kerr black hole, we find the associated central charge used in Cardy formula could be negative if the applied magnetic field on the black hole is superstrong, a condition where the magnetic field parameter $B$ is much larger compared to the inverse of black hole mass $M$. It signals that the dual CFT can be non-unitary, which is not a property of the Kerr-Newman/CFT correspondence \cite{Hartman:2008pb}. Interestingly, when the central charge is negative, the corresponding left temperature $T_L$ (comes from the Boltzmann factor in the theory with a nonvanishing central charge) is also negative, which finally lead to the always positive Cardy entropy. Since we know in thermodynamics that a negative temperature is hotter than a positive one, the result which $T_L$ can be negative is also appealing, i.e. that the dual theory can  really ``mimic'' the very energetic situation of the corresponding black hole. 

The organization of this paper is as follows. In section \ref{s.1}, we review some aspects of magnetized black holes, based on the Melvin-Kerr metric derived in \cite{ernst-wild}. Subsequently, in section \ref{s.2}, we get the near horizon geometry of an extreme Melvin-Kerr black hole, which possesses the $SL(2,\mathbb{R})\times U(1)$ symmetry. Then, in section \ref{s.3} we employ the Asymptotic Symmetry Group (ASG) method \cite{Guica:2008mu} to compute the central charge associated to asymptotic symmetry of near horizon extreme magnetized Kerr (NHEMK) geometry. Section \ref{s.4} is devoted to discuss the Frolov-Thorne temperature in NHEMK, which is then followed by microscopic calculation for the entropy of extreme Melvin-Kerr black holes. In section \ref{s.6}, we discuss briefly the hidden conformal symmetry of a generic Melvin-Kerr black hole. Finally, we give a discussion and our conclusions in section \ref{s.7}. In this paper, we use the units where $c = G = \hbar = 1$. 

\section{Magnetized Kerr spacetime}\label{s.1}

This section presents a review of Melvin-Kerr black hole, i.e. Kerr black hole immersed in a background magnetic field. Quite recently, some studies which discuss several aspects of Melvin-Kerr spacetime and the black hole notion in this geometry appear in literature \cite{Booth:2015,Gibbons:2013yq,Gibbons:2013dna}. The textbook definition of black holes using the asymptotic flatness assumption does not fit the picture of black holes in the Melvin magnetic universe, where the spacetime is not asymptotically flat. Clearly, it leads to such an obscurity to discuss black hole physics in Melvin universe, where the magnetized black hole solution can be obtained by performing a Harrison-type transformation \cite{Harrison} to the known black hole solutions in Einstein-Maxwell theory that preserve the asymptotic flatness. Nevertheless, to model the black holes at the center of our galaxy, we could use the Melvin-Kerr solution. Indeed, the asymptotic flatness condition is violated, but we can preserve the other properties of black holes, e.g. the event horizons, surface gravity, Hawking temperature, etc.
 
The Melvin-Kerr metric was first obtained by Ernst and Wild \cite{ernst-wild}, thus sometimes is also called as the Ernst-Wild solution \cite{Hiscock:1980zf,Aliev:1988,Aliev:1988wy,Aliev:1989wz}. To obtain the Melvin-Kerr metric, one can apply a Harrison-type transformation to the ``Ernst potential'' ${\cal E}$ that belongs to the Kerr solution,
\be \label{E-pot}
{\cal E}' = \frac{\cal E}{\Lambda}\,,
\ee
\be \label{Phi-pot}
\Phi ' = -\frac{{B{\cal E}}}{2\Lambda}\,,
\ee
where $\Lambda  = 1 - {\textstyle{1 \over 4}}B^2 {\cal E}$. The strength of magnetic fields appearing in the spacetime as a result of the magnetizing transformation (\ref{E-pot}) and (\ref{Phi-pot}) is given by the constant parameter $B$. In the case of obtaining the Melvin-Kerr-Newman solution, the corresponding Harisson-type transformation is applied to the Ernst potentials ${\cal E}$ and $\Phi$ of Kerr-Newman solution, namely
\[
{\cal E}' = \frac{{\cal E}}{\Lambda }~~,~~\Phi ' = \frac{1}{\Lambda }\left( {\Phi  - \frac{{B{\cal E}}}{2}} \right)\,,
\]
where $\Lambda  = 1 + B\Phi  - \frac{{B^2 {\cal E}}}{4}$. Note that these ``Ernst potentials'' ${\cal E}$ and $\Phi$ can be defined with respect to the Killing vectors $\partial_t$ or $\partial_\phi$ for a stationary and axially symmetric spacetime. In this paper we adopt the coordinate system adapted to the congruence ${\bf \eta} = \partial_\phi$ \cite{Stephani:2003tm}
\be \label{metric.congruence.phi}
ds^2  = l_{\mu \nu } dx^\mu  dx^\nu   - f\left( {d\phi  + w_\mu  dx^\mu  } \right)^2 \,,
\ee
which also appears in \cite{ernst-wild}. With respect to the Killing vector $\partial_\phi$, one can define the corresponding complex gravitational Ernst potential $\cal E$, where according to (\ref{metric.congruence.phi}) we have ${\mathop{\rm Re}\nolimits} \left( {\cal E} \right) = f$. Hence, the Ernst gravitational and electromagnetic potentials which are discussed in this paper are defined with respect to the Killing vector $\partial_\phi$, including the ones in appendix \ref{app.Ernst} where some details on the potentials ${\cal E}$ and $\Phi$ of Kerr-Newman solution are given. However, since the discussion in this paper is limited to the case of Melvin-Kerr spacetime only, then we restrict the Harisson-type transformation up to the one in (\ref{E-pot}) and (\ref{Phi-pot}). 

Accordingly, the functions $f$ and $w$ which appear in the Ernst potentials 
\be \label{Ernst.Kerr}
{\cal E} = f + i\phi  - \left| \Phi  \right|^2 ~~,~~ \nabla w =  - \frac{\rho }{{f^2 }}\left( {i\nabla \phi  + \Phi ^* \nabla \Phi  - \Phi \nabla \Phi ^* } \right)\,,
\ee
construct the stationary and axial symmetric line element
\be\label{Kerr-Lewis}
ds^2  = f^{ - 1} \left( {\rho ^2 dt^2  - 2P^{ - 2} d\zeta d\zeta ^* } \right) - f\left( {d\phi  - w dt} \right)^2 \,.
\ee 
In the equation above, $\rho$, $P$, and $\zeta$ are some functions that depend on the spacetime coordinates, and $\nabla$ is the gradient operator related to the line element ${d\zeta d\zeta ^* }$. Following Ernst \cite{Ernst}, the ``non-magnetized'' line element (\ref{Kerr-Lewis}) is related to the magnetized one
\be\label{mag.line.element}
ds^2  = \left( {f'} \right)^{ - 1} \left( {\rho ^2 dt^2  - 2P^{ - 2} d\zeta d\zeta ^* } \right) - f'\left( {d\phi  - w'dt} \right)^2 
\ee 
through the relations
\be\label{f-trans}
f' = \left| \Lambda  \right|^2 f\,,
\ee 
and 
\be \label{w-trans}
\nabla w ' = \left| \Lambda  \right|^2 \nabla w  + \rho f^{ - 1} \left( {\Lambda ^* \nabla \Lambda  - \Lambda \nabla \Lambda ^* } \right)\,.
\ee 

The transformations (\ref{f-trans}) and (\ref{w-trans}) yield a spacetime in magnetic universe. Performing these transformations to a solution of the vacuum Einstein equation produces a corresponding ``magnetized'' solution in the Einstein-Maxwell system. As an example, let us consider the four dimensional Minkowski spacetime, 
\be\label{Minkowski.metric}
ds^2  =  - dt^2  + d\rho ^2  + dz^2  + \rho ^2 d\phi ^2 \,
\ee 
which can be obtained from a Kerr black hole solution in the limit of vanishing black hole's mass and angular momentum. The line element (\ref{Minkowski.metric}) can be brought into the form of eq. (\ref{Kerr-Lewis})
by setting \cite{Ernst} $f=-\rho^2$, $P=\rho^{-1}$, $w=0$, and $d\zeta = (dz+id\rho)/\sqrt{2}$. The ``magnetized'' version of (\ref{Minkowski.metric}) due to the transformations (\ref{f-trans}) and (\ref{w-trans}) is known as the Melvin magnetic universe \cite{Ernst},
\be\label{mag-Minkowski}
ds^2  = \Lambda_M ^2 \left( { - dt^2  + d\rho ^2  + dz^2 } \right) + \Lambda_M ^{ - 2} \rho ^2 d\phi ^2 \,,
\ee 
where $\Lambda_M = 1 + {\textstyle{1 \over 4}}B^2 \rho^2$. The metric (\ref{mag-Minkowski}) is not a solution to the vacuum Einstein equation anymore. It rather solves the Einstein-Maxwell equations whose vector field solution gives rise to the Cartan components of the magnetic field which reads \cite{Ernst} $H_i dx^i  = B\Lambda_M^{ - 2} dz$. For the Schwarzschild spacetime, the resulting transformed line element after equations (\ref{E-pot})-(\ref{w-trans}) are employed can be read as
\be\label{metric.mag.sch}
ds^2  = \Lambda_S ^2 \left( { - \left( {1 - \frac{{2M}}{r}} \right)dt^2  + \left( {1 - \frac{{2M}}{r}} \right)^{ - 1} dr^2  + r^2 d\theta ^2 } \right) + \frac{{r^2 \sin ^2 \theta d\phi ^2 }}{{\Lambda_S ^2 }}
\ee 
where $\Lambda_S = 1 + {\textstyle{1 \over 4}}B^2 r^2 \sin^2\theta$. Accordingly, the Cartan components of the magnetic field are \cite{Ernst}
\be
H_i dx^i  = \frac{B}{{\Lambda_S ^2 }}\left( {\cos \theta dr - \sqrt {1 - \frac{{2M}}{r}} \sin \theta d\theta } \right)\,,
\ee 
where $B$ will be called as the magnetic field parameter for the rest of this paper.

In this paper, we will discuss the Melvin-Kerr black hole, which reduces to the Melvin magnetic universe (\ref{mag-Minkowski}) for $M=0$ and $a=0$. The corresponding line element that describes this geometry is given by Ernst and Wild \cite{ernst-wild}, which reads
\be \label{metric-EW}
ds^2  = \Sigma \left| \Lambda  \right|^2 \left( { - \frac{\Delta }{\Xi}d\tilde t^2  + \frac{{d\tilde r^2 }}{{\Delta }} + d\theta^2 } \right) + \frac{{\Xi \sin ^2 \theta }}{{\Sigma \left| \Lambda  \right|^2 }}\left( {d\tilde \phi  - w' d{\tilde t}} \right)^2  \,.
\ee
Metric (\ref{metric-EW}) is obtained by employing the transformations (\ref{E-pot})-(\ref{w-trans}) to the Kerr line element\footnote{The metrics (\ref{metric-EW}) and (\ref{metric-Kerr}) are expressed in the Boyer-Lindquist coordinates $\left( {\tilde t,\tilde r,\theta ,\tilde \phi } \right)$ instead of $\left( { t, r,\theta , \phi } \right)$ as in (\ref{metric.mag.sch}) hence we can use $\left( { t, r,\theta , \phi } \right)$ in writing the near horizon line element (\ref{metric-MKnh}).}
\be \label{metric-Kerr}
ds^2  = \Sigma \left( { - \frac{\Delta }{\Xi}d\tilde t^2  + \frac{{d\tilde r^2 }}{{\Delta }} + d\theta^2 } \right) + \frac{{\Xi \sin ^2 \theta }}{{\Sigma }}\left( {d\tilde \phi  -  w d{\tilde t}} \right)^2  \,,
\ee 
where $\Xi = \left( {\tilde r^2  + a^2 } \right)^2  - \Delta a^2 \sin ^2 \theta $, $\Sigma  = \tilde r^2  + a^2 \cos ^2 \theta $, $\Delta  = \tilde r^2  + a^2  - 2M\tilde r$, and black hole's angular momentum $J = Ma$. In (\ref{metric-EW}), the real and imaginary parts of $\Lambda$ are \cite{Gibbons:2013yq,Aliev:1988}
\be 
{\mathop{\rm Re}\nolimits}~ \Lambda  = 1 + \frac{{B^2 }}{4}\left( {\left( {\tilde r^2  + a^2 } \right)\sin ^2 \theta  + \frac{{2a^2 M\tilde r\sin ^4 \theta }}{\Sigma }} \right)\,,
\ee
and
\be 
{\mathop{\rm Im}\nolimits}~ \Lambda  =  - \frac{{B^2 \cos \theta }}{4}\left( {2aM\left( {2 + \sin ^2 \theta } \right) + \frac{{2a^3 M\sin ^4 \theta }}{\Sigma }} \right)\,,
\ee 
respectively. Moreover, the corresponding $\zeta$, $P$, $\rho$, $w$, and $f$ in the general metric (\ref{Kerr-Lewis}) for Kerr line element (\ref{metric-Kerr}) are \cite{ernst-wild}
\be
d\zeta  = \frac{{dr}}{{\sqrt {2\Delta } }} + i\frac{{d\theta }}{{\sqrt 2 }}~,~\rho  = \frac{{\sin \theta }}{{\sqrt \Delta  }}~,~
P = \left( {\sqrt \Xi  \sin \theta } \right)^{ - 1} ~,~ f =  - \frac{{\Xi \sin ^2 \theta }}{\Sigma } ~,~w = \frac{{2Mra}}{\Xi }\,.
\ee

However, Hiscock found that the magnetized Kerr metric proposed by Ernst and Wild (\ref{metric-EW}) suffers a conical singularity in $\tilde\phi$ coordinate. This can be seen easily from the transformation (\ref{f-trans}) which leads to a one full round in $\tilde\phi '$ is not $2\pi$ anymore, but $2\pi \left| \Lambda_0  \right|^2$ instead \cite{Hiscock:1980zf} where the radius independent $\Lambda_0$ is defined as $\Lambda_0 = \Lambda \left(\theta=0\right)$. Therefore, a scaling in $\tilde\phi ' \to \left| \Lambda_0  \right|^{-2}\tilde\phi '$ is considered as the cure to the problem \cite{Hiscock:1980zf,Aliev:1988wy,Aliev:1989wz}, thence we have now $2\pi$ in one full round. Therefore, in discussing the Kerr black holes immersed by a magnetic field, we use the modified Ernst-Wild metric proposed by Aliev, et al, \cite{Aliev:1988wy,Aliev:1989wz}
\be \label{metric-MK}
ds^2  = \Sigma \left| \Lambda  \right|^2 \left( { - \frac{\Delta }{\Xi}d\tilde t^2  + \frac{{d\tilde r^2 }}{{\Delta }} + d\theta^2 } \right) + \frac{{\Xi \sin ^2 \theta }}{{\Sigma \left| \Lambda  \right|^2 }}\left( {\left| {\Lambda _0 } \right|^2 d\tilde \phi  - w' d{\tilde t}} \right)^2  \,,
\ee
where 
\be\label{wtilde}
w' = \frac{{16M\tilde ra + w_B \left( {\tilde r,\theta } \right)}B^4}{8\Xi}
\ee 
and
\[
w_B \left( {\tilde r,\theta } \right) = 4a^3 m^3 {\tilde r}\left( {3 + \cos ^4 \theta } \right)~~~~~~~~~~~~~~~~~~~~~~~~~~~~~~~~~~~~~~~~~~~~~~~~~~~~~~~~~~~~~~
\]
\[
~~~~~~~~~~~~~~ + 2am^2 \left( {{\tilde r}^4 \left( {\left( {\cos ^2 \theta  - 3} \right)^2  - 6} \right) + 2a^2 {\tilde r}^2 \left( {3 - 3\cos ^2 \theta  - 2\cos ^4 \theta } \right) - a^4 \left( {1 + \cos ^4 \theta } \right)} \right)
\]
\be
~~~~~~~~~~~ + amr\left( {{\tilde r}^2  + a^2 } \right)\left( {{\tilde r}^2 \left( {3 + 6\cos ^2 \theta  - \cos ^4 \theta } \right) - a^2 \left( {1 - 6\cos ^2 \theta  - 3\cos ^4 \theta } \right)} \right)\,.
\ee

It is easy to see that setting $B=0$ in the line element (\ref{metric-MK}) gives us the Kerr metric. Solving $\Delta = 0$ gives the position of black hole horizons, i.e. $\tilde r_ \pm   = M \pm \sqrt {M^2  - a^2 } $, similar to those in the case of Kerr black holes. Accordingly, the extreme condition would also be the same to that in Kerr, i.e. $a=M$. Furthermore, the angular momentum at the horizon and Hawking temperature of Melvin-Kerr black holes can be read as 
\be \label{OmH}
\Omega _H  = \left. {\frac{{\tilde w}}{{\left| {\Lambda \left( {\theta  = 0} \right)} \right|^2 }}} \right|_{r = r_ +  } =\frac{{16M\tilde r_ +  a + \tilde w_B \left( {\tilde r_ +  ,\theta } \right)B^4 }}{{8\left( {\tilde r_ + ^2  + a^2 } \right)^2 }}
\ee
and 
\be \label{TH}
T_H  = \frac{{\hbar \kappa }}{{2\pi }}
\ee
respectively. The surface gravity $\kappa$ of Melvin-Kerr is the same with that of Kerr black holes \cite{Booth:2015},
\be
\kappa  = \frac{{\tilde r_ +   - M}}{{\tilde r_ + ^2  + a^2 }}\,.
\ee 

Together with the Melvin-Kerr metric solution (\ref{metric-MK}), the following vector fields ${\bf A} = A_\mu dx^\mu$ are the fields in Einstein-Maxwell system \cite{Gibbons:2013yq},
\be 
{\bf A} = \left( {\Phi _0  - {w'}\Phi _3 } \right)d{\tilde t} + \Phi _3 d{\tilde \phi} 
\ee
where
\[
\Phi _0  =  - \frac{a}{{8\Xi }}\left\{ {4a^4 M^2  + 2a^4 M} \right.\tilde r - 24a^2 M^3 \tilde r - 24a^2 M^2 \tilde r^2  - 4a^2 M\tilde r^3  - 12M^2 \tilde r^4 - 6M\tilde r^5
\]
\be  
\left. {  - \Delta \left( {12\tilde rM\left( {\tilde r^2  + a^2 } \right)\cos ^2 \theta  + \left( {2M\tilde r^3  + a^2 \left( {4M^2  - 6M\tilde r} \right)} \right)\cos ^4 \theta } \right)} \right\}\,,
\ee
\[
\Phi _3  = \frac{1}{{8\Sigma \left| \Lambda  \right|^2  }}\left\{ {4\Xi B\sin ^2 \theta  + B^4 \left\{ {\Sigma \left( {\tilde r^2  + a^2 } \right)^2 \sin ^4 \theta  + 4a^2 M\tilde r\left( {\tilde r^2  + a^2 } \right)\sin ^6 \theta } \right.} \right.
\]
\be 
\left. {\left. { + 4a^2 M^2 \left( {\tilde r^2 \left( {2 + \sin ^2 \theta } \right)\cos ^2 \theta  + a^2 \left( {1 + \cos ^2 \theta } \right)^2 } \right)} \right\}} \right\}\,,
\ee
and $w'$ is given in (\ref{wtilde}).

\section{Near-horizon geometry of Magnetized Kerr}\label{s.2}

According to the conjectured Kerr/CFT correspondence, a hint that a 2D CFT could be a dual description of extremal rotating black holes comes from the $SL(2,R) \times U(1)$ symmetry which is possessed by the near horizon of an extremal black hole \cite{Guica:2008mu,Hartman:2008pb,Bredberg:2011hp,Compere:2012jk}. Therefore, it is crucial to get the near horizon geometry of an extreme Melvin-Kerr black hole, and show that the geometry has a set of Killing vectors that reflects this symmetry. We find that the following set of coordinate transformations
\be \label{nhtrans1}
\tilde t = \frac{{2M^2\tau }}{\lambda }
~~,~~
\tilde r = \lambda y + M
~~,~~
\tilde \phi  = \varphi  + \frac{{\left( {1 + 2B^4 M^4} \right)M\tau}}{{\left( {1 + B^4 M^4 } \right)\lambda }}
\ee
transforms (\ref{metric-MK}) to the near horizon geometry of an extreme Melvin-Kerr black hole in $(\tau, y, \theta ,\varphi)$ coordinates, where the explicit line element reads
\be \label{metric-nhMKnotglobal}
ds^2  = \Sigma _ +  \left| \Lambda  \right|_ + ^2 \left( { - y^2 d\tau ^2  + \frac{{dy^2 }}{{y^2 }} + d\theta ^2 } \right) + \frac{{A_ +  \sin ^2 \theta \left| {\Lambda _0 } \right|_ + ^2 }}{{\Sigma _ +  \left| \Lambda  \right|_ + ^2 }}\left( {d\varphi  + \frac{{\left( {1 - B^4 M^4 } \right)}}{{\left| {\Lambda _0 } \right|_ + ^2 }}yd\tau} \right)^2 \,.
\ee
The subscript ``$+$'' in equation above denotes that the corresponding variables have been evaluated at the outer horizon radius $r_+$,
\be 
\Sigma _ +   = M^2 \left( {1 + \cos ^2 \theta } \right)
~~,~~
A_ +   = 4M^2 
\ee 
\be
\left| \Lambda  \right|_ + ^2  = \frac{{\left( {1 + B^2 M^2 } \right)^2  + \left( {1 - B^2 M^2 } \right)^2 \cos ^2 \theta }}{{1 + \cos ^2 \theta }}
~~,~~
\left| {\Lambda _0 } \right|_ + ^2  = \left. {\left| \Lambda  \right|_ + ^2 } \right|_{\theta  = 0} 
\ee

To get the global covering of the spacetime (\ref{metric-nhMKnotglobal}), one can employ a set of transformations \cite{Mei:2010wm},
\be \label{y-global}
y = r + \sqrt {1 + r^2 } \cos t \,,
\ee 
\be 
\tau  = \frac{{\sqrt {1 + r^2 } \sin t}}{{r + \sqrt {1 + r^2 } \cos t}} \,,
\ee
\be \label{phi-global}
\varphi  = \phi  + \frac{{\left( {1 - B^4 M^4 } \right)}}{{\left| {\Lambda _0 } \right|_ + ^2 }}\ln \left( {\frac{{1 + \sqrt {1 + r^2 } \sin t}}{{\cos t + r\sin t}}} \right)\,.
\ee
The transformed version of the metric (\ref{metric-nhMKnotglobal}) due to the mappings (\ref{y-global}) - (\ref{phi-global}) is
\be \label{metric-MKnh}
ds^2  = \Sigma _ +  \left| \Lambda  \right|_ + ^2 \left( { - \left( {1 + r^2 } \right)dt^2  + \frac{{dr^2 }}{{\left( {1 + r^2 } \right)}} + d\theta ^2 } \right) + \frac{{A_ +  \sin ^2 \theta \left| {\Lambda _0 } \right|_ + ^2 }}{{\Sigma _ +  \left| \Lambda  \right|_ + ^2 }}\left( {d\phi  + \frac{{\left( {1 - B^4 M^4 } \right)}}{{\left| {\Lambda _0 } \right|_ + ^2 }}rdt} \right)^2 \,.
\ee
Turning the magnetic field parameter $B$ off from the last equation, we get the near horizon geometry for an extreme Kerr black holes which appears in the paper by Guica, et al. \cite{Guica:2008mu}. It is easy to see that $\partial_\phi$ is a Killing vector of (\ref{metric-MKnh}) and generates the $U(1)$ symmetry of spacetime. In addition to this $U(1)$ generator, the following generators of $SL(2,{\mathbb{R}})$ group
\be \label{J0}
\tilde J_0  = 2\partial _t \,,
\ee
\be \label{J1}
\tilde J_1  = \frac{{2r\sin t}}{{\sqrt {1 + r^2 } }}\partial _t  - 2\cos t\sqrt {1 + r^2 } \partial _r  + \frac{{2\sin t}}{{\sqrt {1 + r^2 } }}\partial _\phi  \,,
\ee
\be \label{J2}
\tilde J_2  =  - \frac{{2r\cos t}}{{\sqrt {1 + r^2 } }}\partial _t  - 2\sin t\sqrt {1 + r^2 } \partial _r  - \frac{{2\cos t}}{{\sqrt {1 + r^2 } }}\partial _\phi  \,,
\ee
are also the Killing vectors of (\ref{metric-MKnh}) up to a rescaling\footnote{See appendix \ref{KV} for a discussion.} of $\phi$. At this point we have seen that the near horizon geometry of an extreme Melvin-Kerr black hole (\ref{metric-MKnh}) possesses the $SL(2,{\mathbb R})\times U(1)$ symmetry which indicates the possibility to apply the ASG method in getting a central charge associated to the asymptotic symmetry of the spacetime.

\section{Central Charge}\label{s.3}

In this section, adopting the ASG method by Guica, et al. \cite{Guica:2008mu}, we compute the central charge that appear in the Virasoro algebra between conserved charges $Q_\zeta$ related to the symmetry transformation (Lie derivative) ${\cal L}_\zeta$ of fields in the corresponding Einstein-Maxwell system.  These Lie derivatives, i.e. ${\cal L}_\zeta g_{\mu\nu}$ and ${\cal L}_\zeta A_\mu$, subject to some appropriate boundary conditions. For NHEMK spacetime (\ref{metric-MKnh}), the same boundary condition at asymptotic radius $r$ as the one applies to NHEK \cite{Guica:2008mu,Hartman:2008pb} can also be employed, i.e.
\be\label{hmn}
{\cal L}_\zeta g_{\mu\nu} = h_{\mu\nu}  \sim \left( {\begin{array}{*{20}c}
   {{\cal O}\left( {r^2 } \right)} & {{\cal O}\left( 1 \right)} & {{\cal O}\left( {r^{ - 1} } \right)} & {{\cal O}\left( {r^{ - 2} } \right)}  \\
   {{\cal O}\left( 1 \right)} & {{\cal O}\left( 1 \right)} & {{\cal O}\left( {r^{ - 1} } \right)} & {{\cal O}\left( {r^{ - 1} } \right)}  \\
   {{\cal O}\left( {r^{ - 1} } \right)} & {{\cal O}\left( {r^{ - 1} } \right)} & {{\cal O}\left( {r^{ - 1} } \right)} & {{\cal O}\left( {r^{ - 2} } \right)}  \\
   {{\cal O}\left( {r^{ - 2} } \right)} & {{\cal O}\left( {r^{ - 1} } \right)} & {{\cal O}\left( {r^{ - 2} } \right)} & {{\cal O}\left( {r^{ - 3} } \right)}  \\
\end{array}} \right) \,.
\ee
This $h_{\mu\nu}$ can be considered as the deviation of NHEMK metric $g_{\mu\nu}$ in (\ref{metric-MKnh}), while in the computation we perform the lowering/raising tensorial indices as well as covariant derivatives by using $g_{\mu\nu}$. Accordingly, the general form of diffeomorphisms $\zeta$ which obey the boundary condition (\ref{hmn}) is \cite{Guica:2008mu}
\be \label{zeta}
\zeta ^\mu  \partial _\mu   =  - r\varepsilon '\left( \phi  \right)\partial _r  + \varepsilon \left( \phi  \right)\partial _\phi\,,
\ee  
where $(~)'$ stands for the differentiation with respect to $\phi$. 

It is shown in appendix \ref{app.nh-vector} that the near horizon vector field solution in Einstein-Maxwell equation to the spacetime (\ref{metric-nhMKnotglobal}) has a general form as the one discussed in \cite{Hartman:2008pb}
\be\label{Afrt}
A_\mu  dx^\mu   = f\left( \theta  \right)\left( {krtdt + d\phi } \right)\,.
\ee 
The authors of \cite{Hartman:2008pb} found that, after a boundary condition to the vector field at a large $r$
\be a_\mu   \sim \left( {{\cal O}\left( r \right),{\cal O}\left( {r^{ - 1} } \right),{\cal O}\left( 1 \right),{\cal O}\left( {r^{ - 2} } \right)} \right)\ee 
is imposed, the gauge + diffeomorphism transformation to the kind of vector field in (\ref{Afrt}) will not contribute to the total central charge. Consequently, in the case of NHEMK, the total central charge due to the diffeomorphism transformation of the fields comes from the gravity only, as in the case of near horizon extremal Kerr-Newman black holes \cite{Hartman:2008pb}. Hence, in the following Poisson bracket computation of conserved charges below, we will consider the central term which is a result of the spacetime diffeomorphism only.

Under the diffeomorphism $\zeta$, in regard to the Einstein-Hilbert 4-form Lagrangian\footnote{The complete Lagrangian would be the Einstein-Maxwell one. However, the gauge transformation and diffeomorphism of $A_\mu$ will not contribute to the central term, then we can simply perform in quite detail the calculations based on Einstein-Hilbert Lagrangian only.}
\be 
{\rm {\bf L}}=\frac{R}{{16\pi }} * {\rm {\bf 1}}\,,
\ee 
one can find the corresponding 3-form Noether current \cite{Mei:2010wm} 
\be
{\rm {\bf J}}_\zeta   =  - \frac{1}{{16\pi }}\left( {d^3 x} \right)_\mu  \nabla _\nu  \left( {\nabla ^\mu  \zeta ^\nu   - \nabla ^\nu  \zeta ^\mu  } \right)\,,
\ee 
and the associated 2-form conserved current
\be
{\rm {\bf Q}}_\zeta   =  - \frac{1}{16\pi }\left( {d^2 x} \right)_{\mu \nu}  \left( {\nabla ^\mu  \zeta ^\nu   - \nabla ^\nu  \zeta ^\mu  } \right)\,.
\ee 
In the last two equations we have used the notation
\be 
\left( {dx^{n - p} } \right)_{\mu _1 ...\mu _p }  = \frac{1}{{p!\left( {n - p} \right)!}}\varepsilon _{\mu _1 ...\mu _p \nu _1 ...\nu _{n - p} } dx^{\nu _1 }  \wedge ... \wedge dx^{\nu _{n - p} } \,,
\ee 
where $\left| {\varepsilon _{...} } \right| = \sqrt {\left| g \right|} $. 

Defining
\be
Q_\zeta   = \int_{\cal V} {d{\rm {\bf Q}}_\zeta  }  = \oint_{\partial {\cal V}} {{\rm {\bf Q}}_\zeta  } 
\ee 
where ${\cal V}$ and $\partial {\cal V}$ are the space-like slice and its boundary of NHEMK spacetime respectively. The Poisson bracket between two $Q_{\zeta}$'s can be found to be \cite{Mei:2010wm}
\be\label{Q-alg}
\left\{ {Q_\zeta  ,Q_\xi  } \right\} = Q_{\left[ {\zeta ,\xi } \right]}  + C\left[ {\zeta ,\xi } \right]\,,
\ee 
where the corresponding central term $C\left[ {\zeta ,\xi } \right]$ can be written as
\be \label{central-term}
C\left[ {\zeta ,\xi } \right] =  - \oint_{\partial V} {{\rm{{\bf k}}}^g_\zeta  \left( {g_{\mu\nu} ,{\cal L}_\xi  g_{\mu\nu} } \right)}\,.
\ee
Explicitly, the two form ${{\rm{{\bf k}}}^g_\zeta  \left( {g_{\mu\nu} ,h_{\mu\nu} } \right)}$ in central term (\ref{central-term}) is given by
\be \label{k_mn}
{{\rm{{\bf k}}}^g_\zeta  \left( {g_{\mu\nu} ,h_{\mu\nu} } \right)} = \frac{{\sqrt {\left| g \right|} }}{64\pi}\varepsilon _{\mu \nu \alpha \beta } k^{\mu\nu}_\zeta dx^\alpha   \wedge dx^\beta
\ee 
where\footnote{$h_{\mu \nu }  = \nabla _\mu  \zeta _\nu   + \nabla _\nu  \zeta _\mu $ where $\nabla_\mu$ is the covariant derivative related to the metric tensor (\ref{metric-MKnh}). Raising and lowering index are performed by using $g_{\mu\nu}$.}
\be 
k^{\mu\nu}_\zeta =\zeta^\nu\nabla^\mu h - \zeta^\nu \nabla_\rho h^{\mu\rho}
+\frac{h}{2}\nabla^{ \nu}\zeta^{\mu} -h^{\nu\rho}\nabla_\rho
\zeta^\mu+\zeta_\rho\nabla^{\nu}h^{\mu\rho} - (\mu \to \nu)\,.
\ee 

The corresponding $k^{\mu\nu}_\zeta$ that contributes to the computation of central term (\ref{central-term}) associated to the diffeomorphism (\ref{zeta}) is
\be 
k^{tr}_\zeta  = \frac{{in^2 \left( {m - n} \right)\left( {1 - B^4 M^4 } \right)e^{ - i\left( {m + n} \right)\phi } }}{{M^2 \left( {\left( {1 + B^2 M^2 } \right) + \left( {1 - B^2 M^2 } \right)\cos ^2 \theta } \right)\left( {1 + B^4 M^4 } \right)}}\,.
\ee 
In the last equation we only consider the terms that couple to $m^3$ after $\delta_{m+n}$ is employed, since only this term matters to indicate the central charge associated to the algebra between asymptotic symmetry charges $Q_\xi$. Finally the central term for NHEMK can be shown to be 
\be
K\left[ {\zeta _m ,\zeta _n } \right] =  - \frac{{i\left( {m - n} \right)n^2 \pi M^2 \left( {1 - B^4 M^4 } \right)}}{{2\pi }}\delta _{m + n} \,,
\ee 
which leads to the central charge \cite{Mei:2010wm}
\be \label{c-nhemk}
c_{grav} = 12M^2 \left(1-B^4 M^4 \right)\,.
\ee 
In the absence of magnetic field parameter $B$, the central charge (\ref{c-nhemk}) reduces to the one in non-magnetized extreme Kerr case \cite{Guica:2008mu}.

Unlike the central charge in the case of extremal Kerr-Newman/CFT correspondence \cite{Hartman:2008pb}, which is always positive, the central charge associated to the near horizon of extremal Melvin-Kerr black hole (\ref{c-nhemk}) is negative when $|B|>1/M$. It is well known that a CFT with negative central charge is non-unitary \cite{CFTbook-Mathieu}, for example the Lee-Yang edge singularity \cite{Lee-Yang} which occurs in the two dimensional Ising model with imaginary magnetic field. This model was studied quite extensively, for example the relation between this singularity with $\Phi^3$ theory was performed in \cite{Fisher:1978pf}, and Cardy in \cite{Cardy:1985yy-Yang-Lee} showed some relations between a two dimensional conformal invariance field theory with this Yang-Lee edge singularity. Even though a non-unitary CFT is quite often to be considered as a ``sick'' theory, it could have some physical applications. For example, it might have some applications in describing conformal turbulence \cite{Polyakov-confturb}.

Related to the central charge (\ref{c-nhemk}), one can interpret that the unitarity of dual 2D CFT can be broken as the magnetic parameter $B$ increases. This is a new feature of the conjecture extremal-rotating/CFT correspondence, where the gravity side is Einstein-Maxwell theory, which does not appear in the case of Kerr-Newman/CFT holography. Moreover, related to the black hole problems, it is interesting to note that the gravity/non-unitary CFT  correspondence is worth some further studies. In \cite{Vafa:2014iua}, Vafa pointed out that it could holographically deal with the non-unitarity process in information evolution related to the Hawking radiation. In the present paper, we are not dealing with the problem of non-unitarity which appears in the Hawking radiation, but instead on a duality between some physical aspects of extremal Melvin-Kerr black holes with a non-unitarity CFT.

\section{Frolov-Thorne Temperature}\label{s.4}

It is known that, based on the work by Hartle and Hawking \cite{Hartle:1976tp}, the spacetime outside of the Schwarzschild black hole is populated by quantum fields in the thermal state weighted by a Boltzmann factor  \be e^{ - \omega /T_H }\,.\ee On the other hand, the near horizon of Kerr black holes, the corresponding thermal state is weighted by the Boltzmann factor \be\label{BF-Kerr} e^{ - \left( {\omega  - m\Omega _H } \right)/T_H} \,.\ee  
These Boltzmann factors are obtained by tracing out the parts of density matrix for vacuum\footnote{Hartle-Hawking vacuum for Schwarzschild black hole and Frolov-Thorne one for Kerr.} that belongs to the region inside of the horizon. However, when a Kerr black hole are in extremal state, $T_H$ vanishes, which yields the vanishing Boltzmann factor $e^{ - \left( {\omega  - m\Omega _H } \right)/T_H} $ except in the superradiant condition denoted by $\omega = m\Omega_H$ where $\Omega_H$ is the angular velocity of black holes at the outer horizon. Hence, in the limits $T_H \to 0$ and $\omega\to m\Omega_H$, the corresponding Boltzmann factor can be shown to be $e^{-m/T_L}$, where 
\be \label{FT-temp}
T_L = \frac{1}{2\pi}
\ee 
which is known as the extremal Frolov-Thorne temperature \cite{Frolov-Thorne,Compere:2012jk}. In the context of 2D CFT dual, $T_L$ is known as the left-mover temperature, obtained by rewriting the Boltzman factor (\ref{BF-Kerr}) as a product of Boltzmann factors from the left and right mover dual theories. The near horizon of an extreme Kerr black hole is populated by quantum fields in the superradiant modes $\omega = m\Omega_H$ and thermal state with temperature $T_L$. To obtain the temperature (\ref{FT-temp}), we have used the expansion for a scalar fields as
\be\label{scalar-exp-Kerr}
\phi  = \sum\limits_{lm\omega } {C_{lm\omega } e^{ - i\omega \tilde t + im\tilde \phi } f_l \left( {{\tilde r},\theta } \right)} \,.
\ee 

Since the Melvin-Kerr spacetime is stationary and possesses the axial symmetry, then the expansion (\ref{scalar-exp-Kerr}) still applies in that geometry. Following \cite{Guica:2008mu}, the associated left-moving and right-moving temperatures in the quantum theory with the Frolov-Thorne vacuum in extreme Melvin-Kerr spacetime can be obtained as follows. By using the coordinate transformation (\ref{nhtrans1}), one can obtain
\be \label{e-rel-1}
e^{ - i\omega \tilde t + im\tilde \phi }  = e^{ - \frac{i}{\lambda }\left( {2M^2 \omega  - \frac{{mM\left( {1 + 2B^4 M^4 } \right)}}{{\left( {1 + B^4 M^4 } \right)}}} \right) t  + im\phi }  \equiv  e^{ - in_R t  + in_L \phi } \,.
\ee
From the last equation we can get
\be 
n_L = m~~~~,~~~~n_R  = \frac{M}{\lambda }\left( {2M\omega  - \frac{{m\left( {1 + 2B^4 M^4 } \right)}}{{\left( {1 + B^4 M^4 } \right)}}} \right)\,,
\ee
which are, respectively, the left and right charges associated to the Killing vectors $\partial_\phi$ and $\partial_t$ in the near region of Melvin-Kerr black hole. In terms of these charges, the Boltzmann factor which corresponds to the vacuum outside of an extremal Melvin-Kerr black hole can be re-expressed as a product of Boltzmann factors coming from the left and right movers theories, namely
\be \label{Boltzmann-nRnL}
e^{ - {{\left( {\omega  - \Omega _H m} \right)} \mathord{\left/
 {\vphantom {{\left( {\omega  - \Omega _H m} \right)} {T_H }}} \right.
 \kern-\nulldelimiterspace} {T_H }}} = e^{-n_L/T_L-n_R/T_R}\,,
\ee
where \footnote{In extreme Melvin-Kerr condition, i.e. $a = M$ and $r_+ = M$.}
\be \label{TrTl}
T_R = 0~~~~,~~~~
T_L  = \frac{{1 + B^4 M^4 }}{{2\pi \left( {1 - B^4 M^4 } \right)}}\,.
\ee

In the absence of magnetic field parameter $B$, the left temperature above becomes $1/2\pi$ as it is expected, i.e. the left temperature of the dual CFT for Kerr black holes \cite{Guica:2008mu}. Interestingly, the left temperature $T_L$ in (\ref{TrTl}) diverges at $|B|\to 1/M$, and even can have negative value when $|B| > M$ even though the Hawking temperature $T_H$ of black holes vanishes at extremality. Note that the increasing of magnetic parameter $B$ does not lead to the violation of black hole censorship conjecture, unlike the Kerr-Newman black holes that can be broken by allowing a test body capture which yield to the violation of extremal bound $M^2>a^2+Q^2$ \cite{Gao:2012ca}. The situation where temperature can be negative is also found in the nuclear spin systems, first experimentally observed by Purcell and Pound \cite{Purcell-Pound}. Later on, the theoretical studies of thermodynamics with negative temperature were carried out by Ramsey \cite{Ramsey}, where he showed that statistical mechanics as we know nowadays is compatible with the system which has negative absolute temperature \footnote{In Appendix \ref{app.NegativeTemp}, a brief review on the negative temperature is given.}.

The experimental and theoretical studies of nuclear spin systems at negative temperature \cite{Purcell-Pound,Ramsey,Negative-Temp} allow us to interpret that when $T_L < 0$, the higher energy states in the left mover dual 2D CFT are more occupied compared to the ones with lower energy. Moreover, the corresponding dual theory also has an upper bound of energy \cite{Negative-Temp}, so the negative temperature may appear. Nevertheless, finding a unique 2D CFT that is dual to the physics of an extremal Kerr black holes is still in the list of open problems in the Kerr/CFT correspondence until today. Therefore, we are unable to describe more on some aspects of the dual field theory which holographically describes the extremal Melvin-Kerr black holes, especially the corresponding Hamiltonian which allows $T_L <0$.

\section{Microscopic Entropy}\label{s.5}

According to the conjectured Kerr/CFT correspondence, or in general the rotating black holes/CFT duality, the Cardy formula for entropy in 2D CFT is given by
\be
S_{CFT} = \frac{{\pi ^2 }}{3}cT\,,
\ee 
where $T$ is the corresponding Frolov-Thorne temperature can reproduce the Bekenstein-Hawking entropy of the extreme (rotating/charged) black hole under consideration. Since in the previous section only the left mover temperature that is non zero, then the Cardy formula $S_{CFT} = \pi ^2 cT_L /3$ gives the microscopic entropy for an extreme Melvin-Kerr black hole as
\be \label{S-cft}
S_{CFT} = 2\pi M^2 \left( {1 + B^4 M^4 } \right)\,,
\ee 
where we have used the central charge (\ref{c-nhemk}) and left temperature in (\ref{TrTl}).

In the macroscopic side, from the area of an extreme Melvin-Kerr black hole's event horizon \cite{Booth:2015}
\be\label{Aext} 
{\cal A}_{\rm ext.} = 8\pi M^2 \left( {1 + B^4 M^4 } \right)\,,
\ee
one can obtain the Bekenstein-Hawking entropy for an extreme Melvin-Kerr black hole which reads
\be \label{S-BH-macro}
S_{\rm BH}  = 2\pi M^2 \left( {1 + B^4 M^4 } \right)\,.
\ee
It can be seen that the microscopic entropy (\ref{S-cft}) matches the macroscopic one (\ref{S-BH-macro}), which allow us to claim that the extremal Kerr/CFT correspondence is still valid in the case of magnetized Kerr black holes. Nevertheless, if such a superstrong magnetic field which yield $T_L <0$ can really exist, then the Kerr/CFT correspondence with an external magnetic field applied to the black hole has a richer properties compared to the neutral Kerr/CFT duality \cite{Guica:2008mu}, especially on the unitary to non-unitary transition of the dual 2D CFT. 

\section{On the Hidden Conformal Symmetries}\label{s.6}

In the dictionary of Kerr/CFT correspondence, the duality between 2D CFT and physics of Kerr black holes is not limited to the extremal case only, but also in the non-extremal condition. However, the near horizon of non-extreme Kerr black holes does not have the warped and twisted product of $AdS_2 \times S^2$ structure, therefore it lacks the $SL(2,{\mathbb R}) \times U(1)$ isometry group which furthermore hinders the application of ASG method \cite{Guica:2008mu} to the non-extreme Kerr black holes. Nonetheless, it was found by Castro, et al. \cite{castro} that the conformal symmetry appears in the radial part of the low frequency massless scalar field equation in the near region of black holes. It was shown by demonstrating that the Laplacian of radial wave function can be rewritten as the squared Casimir operator of $SL(2,{\mathbb R})_L \times SL(2,{\mathbb R})_R$. In the non-extremal case, besides the matching of microscopic and macroscopic entropy\footnote{Again by using Cardy formula by an assumption that the correspondng central charge is invariant to that in the extreme case.}, the duality is established by showing that the scattering amplitudes in the near region of Kerr black holes agree with those of a finite temperature dual 2D CFT. This discussion of the non-extremal Kerr/CFT correspondence is also extended to many cases of rotating and charged black holes \cite{hiddenkerrcft}.

However, the separability of Klein-Gordon equation of massless scalar fields in the spacetime background is crucial in showing the hidden conformal symmetry of the black holes \cite{castro,Bredberg:2011hp}. This separability can be shown using the Killing-Yano tensor \cite{Bredberg:2011hp}, or another related tensor namely the Killing-Stackel tensor \cite{keeler}. Kerr spacetime itself possesses the Killing-Stackel tensor \cite{keeler}, hence it is obvious why Teukolsky managed to separate the Klein-Gordon equation in the Kerr spacetime \cite{Teukolsky:1973ha} into the radial and angular equations. However, without any restrictions to the strength of external magnetic fields that interact with a Kerr black hole, one cannot find a Killing-Stackel tensor for the metric (\ref{metric-MK}), meaning the separation of Klein-Gordon equation in magnetized Kerr spacetime is impossible. 

In order to get the separation between the radial and angular parts of the massless scalar field equation $\nabla_\mu\nabla^\mu \Phi =0$ in the background metric (\ref{metric-MK}), where we use the ansatz $\Phi \sim e^{-i\omega {\tilde t}+im{\tilde \phi}} R({\tilde r})S(\theta)$, the magnetic field parameter must be very weak, i.e. $B\ll 1/M$, and the black hole is slowly rotating as well, i.e. $a\ll M$ \cite{Konoplya:2008hj}. In such considerations, the terms $a^2 B^2$, $B^4$, and the higher order of $B$ can be neglected. The corresponding radial equation to the scalar field then can be written as \cite{Konoplya:2008hj}
\be\label{rad.eq.MK}
\partial _{\tilde r} \left( {\Delta \partial _{\tilde r} R\left( {\tilde r} \right)} \right) + \left( {\frac{{\omega ^2 \left( {{\tilde r}^2  + a^2 } \right)^2  - 4aM{\tilde r}m\omega  + a^2 m^2 }}{\Delta } - B^2 m^2 {\tilde r}^2  + a^2 \omega ^2  + \lambda } \right)R\left( {\tilde r} \right) = 0\,,
\ee 
which, in the absence of magnetic field parameter $B$, reduces to the scalar wave equation in Kerr spacetime
\be \label{rad.eq.Kerr}
\partial _{\tilde r} \left( {\Delta \partial _{\tilde r} R\left( {\tilde r} \right)} \right) + \left( {\frac{{\left( {2M\omega {\tilde r}_ +   - am} \right)^2 }}{{\left( {{\tilde r} - {\tilde r}_ +  } \right)\left( {{\tilde r}_ +   - {\tilde r}_ -  } \right)}} - \frac{{\left( {2M\omega {\tilde r}_ -   - am} \right)^2 }}{{\left( {{\tilde r} - {\tilde r}_ -  } \right)\left( {{\tilde r}_ +   - {\tilde r}_ -  } \right)}} + f\left( {{\tilde r},\omega ,M} \right) - \lambda } \right)R\left( {\tilde r} \right) = 0
\ee
with 
\be \label{f.in.eqtn}f\left( {{\tilde r},\omega ,M} \right) = \left( {{\tilde r}^2  + 2M\left( {2M + {\tilde r}} \right)} \right)\omega ^2\,.
\ee 
In \cite{castro}, it was showed that in the low frequency limit $\omega M \ll 1$, the radial wave equation (\ref{rad.eq.Kerr}) possesses the $SL(2,{\mathbb R})_L \times SL(2,{\mathbb R})_R$ hidden symmetry. The limit $\omega M \ll 1$ is important in obtaining the radial wave equation with this symmetry, since this limit allows one to neglect the term $f\left( {{\tilde r},\omega ,M} \right)$ in the equation (\ref{rad.eq.Kerr}). 

Related to the radial equation (\ref{rad.eq.MK}), one can also apply some limits which finally lead to a radial equation with $SL(2,{\mathbb R})_L \times SL(2,{\mathbb R})_R$ symmetry. Together with the low frequency limit $\omega M \ll 1$, we should also impose the condition where $Bm \ll 1/M$ which agrees to the initial condition where the magnetic field parameter must be weak in order to get a separability for the corresponding Klein-Gordon equation. The latter condition give an extra constrain on the quantum magnetic number $m$, which does not appear in the case of showing the hidden conformal symmetry of a generic Kerr spacetime \cite{castro}. In the case of magnetized Kerr, there is a range of scalar modes related to the quantum number $m$, in order to show the corresponding hidden conformal symmetry $SL(2,{\mathbb R})_L \times SL(2,{\mathbb R})_R$ in the associated radial equation. Nevertheless, after applying the conditions $Bm \ll 1/M$ and low frequency of the scalar field $\Phi$, eq. (\ref{rad.eq.MK}) reduces to (\ref{rad.eq.Kerr}) with the term $f({\tilde r},\omega,M)$ is neglected, since the inner and outer horizon's radii of magnetized Kerr black holes are just those of Kerr black holes. In other words, after employing some conditions to get the hidden conformal symmetry for a non-extremal magnetized Kerr black hole, the corresponding variables in the conformal symmetry generators has no differences to the ones in Kerr black hole's case.

\section{Conclusion}\label{s.7}

In this paper, we have computed the entropy of extreme Melvin-Kerr black holes by using Cardy formula in the framework of Kerr/CFT correspondence. By following the Kerr/CFT prescription \cite{Guica:2008mu,Compere:2012jk}, first we show that the near horizon geometry of Melvin-Kerr black holes has the $SL(2,{\mathbb R}) \times U(1)$ isometry group. This allows us to use of ASG method in obtaining the associated central charge. We find that the obtained microscopic entropy agrees to the macroscopic one, from which we learn that Kerr/CFT correspondence fits in the magnetized Kerr case. However, it is found that to extend the holography to a generic Melvin-Kerr black hole, the magnetic field parameter $B$ must be very weak, which finally results the hidden conformal symmetry generators which take the form of those in non-extremal Kerr/CFT correspondence \cite{castro}. 

However, we find several new interesting features in this paper which do not appear in the Kerr-Newman/CFT discussions \cite{Compere:2012jk}. As the magnetic field parameter $B$ increases, we observe an unitary to non-unitary transition in the conjectured dual 2D CFT theory. In the non-unitary regime of the corresponding dual 2D CFT theory, the associated left mover temperature $T_L$ is negative, meaning that the system is very high energetic. Also quite interesting, there is a point where the symmetry algebra of the conjectured dual 2D CFT is centerless, provided by the vanishing central charge when $BM=1$. We also find that in the limit of weak magnetic field, the corresponding near region and low frequency radial wave equation is similar to that in the case of Kerr black hole. Consequently, one can show the hidden conformal symmetry of the non-extremal Melvin-Kerr black hole, by considering the weak magnetic field limit in addition to near region and low frequency conditions for the test scalar field. Increasing the magnetic field breaks this hidden conformal symmetry, since the corresponding Klein-Gordon equation cannot be decoupled anymore.

As the future projects, we find several interesting works that can be pursued. Further studies on more details of the case $BM\ge 1$ in both gravitational and dual 2D CFT theories are challenging. In the case $BM=1$, the central charge is zero without leading the vanishing of dual entropy, since the associated left mover temperature in this case is singular. This finding is in contrast to the vanishing central charge in Kerr/CFT holography due to the absence of black hole's rotation\footnote{In the Kerr/CFT correspondence \cite{Guica:2008mu}, $c_L = 12J$ and $T_L = 1/2\pi$. Therefore, when $a=0$, thence $S_{Cardy} = 0$.} On the other hand, the case of $BM>1$ requires a non-unitary 2D CFT to be dual theory. Exploring the possibility of non-unitary processes, for example the scattering process around the extremal Melvin-Kerr black holes, which are dual to a non-unitary 2D CFT would be an interesting investigation. Testing the Kerr/CFT correspondence to a black hole solution, which is not in the family of Einstein-Maxwell theory, embedded in a magnetic universe can also be a good project to be done.

 \section*{Acknowledgement}
 
 I thank my colleagues at UNPAR for their support and Reinard Primulando for reading this manuscript. Also, I would like to thank the anonymous reviewers for their comments and suggestions.

\appendix
\section{Killing vectors}\label{KV}
It can be verified that for a general spacetime metric
\[
ds^2  = C\left( \theta  \right)\left( { - \left( {1 + r^2 } \right)dt^2  + \frac{{dr^2 }}{{\left( {1 + r^2 } \right)}} + d\theta ^2 } \right) + \frac{{D\left( \theta  \right)}}{{C\left( \theta  \right)}}\left( {d\phi  + krdt} \right)^2 
\]
where $C\left( \theta  \right)$ and $D\left( \theta  \right)$ are some general $\theta$-dependent functions, the symmetry of spacetime is described by the Killing vectors (\ref{J0}) - (\ref{J2}) , for $k=1$. When $k\ne 1$, the angular coordinate $\phi$ can be scaled, hence the factor $k$ can be absorbed to $D(\theta)$, for example.

\section{Central charge from stretched horizon technique}

In \cite{Carlip:2011ax}, Carlip showed an alternative computation of a central charge associated to the near horizon of extremal Kerr black hole, namely the stretched horizon technique. Subsequently, the authors of \cite{Chen:2011wm} generalized the method and derived a formula for central charge obtained from a metric of general rotating black hole,
\be \label{eq.app.3.1}
ds^2  =  - N^2 d{\tilde t}^2  + h_{{\tilde r}{\tilde r}} d{\tilde r}^2  + h_{\theta \theta } d\theta ^2  + h_{{\tilde \phi} {\tilde \phi} } \left( {d{\tilde \phi}  + N^{\tilde \phi}  d{\tilde t}} \right)^2\,.
\ee

A central charge associated to the metric (\ref{eq.app.3.1}) is given by a general formula
\be
c = \frac{{3fA_{BH} }}{{2\pi }}\,,
\ee 
where $A_{BH}$ is black hole's horizon area and $f$ is some functions related to the corresponding spacetime metric, i.e.
\be\label{fgen}
f = \left. {\frac{{f_2 f_3 }}{{f_1 }}} \right|_{{\tilde r} = {\tilde r}_ +  } \,,
\ee 
\be \
f_1 = \frac{N}{({\tilde r}-{\tilde r}_+)}~~~,~~~f_2 = \sqrt{({\tilde r}-{\tilde r}_+)h_{{\tilde r}{\tilde r}}}~~~,~~~f_3 = \frac{N^{\tilde \phi} + \Omega_H}{({\tilde r}-{\tilde r}_+)}\,.
\ee 

Therefore, from the Melvin-Kerr line element (\ref{metric-MK}), we have
\be \label{f123-MK}
f_1  = \sqrt {\frac{{\Sigma \left| \Lambda  \right|^2 }}{A}} ~~~,~~~f_2  = \sqrt {\Sigma \left| \Lambda  \right|^2 } ~~~,~~~f_3  = \frac{{\Omega _H \left| {\Lambda _0 } \right|^2  - w'}}{{\left( {{\tilde r} - {\tilde r}_ +  } \right)\left| \Lambda  \right|^2 }}\,.
\ee
Plugging the functions in (\ref{f123-MK}) into (\ref{fgen}) gives us the desired result (\ref{c-grav}).

\section{Near horizon MK as Einstein-Maxwell system}
\label{app.nh-vector}
In this appendix we show how the get the gauge field solutions in Einstein-Maxwell equations,
\be \label{EMeqtn}
R_{\mu\nu}=T_{\mu\nu}
\ee 
where the energy momentum tensor is
\be 
T_{\mu\nu}=F_{\mu \alpha } F_\nu ^\alpha   - \frac{1}{4}g_{\mu \nu } F_{\alpha \beta } F^{\alpha \beta } 
\ee 
for the near horizon MK geometry. In order to simplify the computation, we introduce a new coordinate $x=\cos\theta$, and work in the metric where the scaling of $\phi$ to avoid the conical singularity has not been considered\footnote{To get the solution in coordinate where the conical singularity has been removed, one can perform the tensor transformation due to the coordinate changes.}. In this consideration, the near horizon geometry of MK can be read as
\be 
ds^2  = \Gamma \left( x \right)\left[ { - r^2 dt^2  + \frac{{dr^2 }}{{r^2 }} + d\theta ^2  + \gamma \left( x \right)\left\{ {d\phi  + krdt} \right\}^2 } \right]\,,
\ee
where
\be 
\Gamma \left( x \right) = M^2 \left( {{\cal Q}^2  + {\cal P}^2 x^2 } \right)\,,
\ee
\be
\gamma \left( x \right) = \frac{{2\left( {1 - x^2 } \right)^{1/2} }}{{\left( {{\cal Q}^2  + {\cal P}^2 x^2 } \right)}}\,,
\ee 
and $k = -{\cal QP}$. In the equations above, we have used new notations, i.e. ${\cal Q} =1 + B^2 M^2$ and ${\cal P} = 1- B^2 M^2$. 

By using the ansatz for vector fields $A_\mu dx^\mu = -{\cal PQ}  L(x) r dt + L(x) d\phi $, the Maxwell equation 
\be 
\nabla _\mu  F^{\mu \nu }  = 0
\ee
reads
\be \label{MeqtnL}
\left( {{\cal P}^2 x^2  + {\cal Q}^2 } \right)^2 \frac{{d^2 L\left( x \right)}}{{dx^2 }} + 2x{\cal P}^2 \left( {{\cal P}^2 x^2  + {\cal Q}^2 } \right)\frac{{dL\left( x \right)}}{{dx}} + 4{\cal P}^2 {\cal Q}^2 L\left( x \right) = 0
\ee 
In the other hand, the Einstein-Maxwell equations (\ref{EMeqtn}) gives
\be \label{EMeqtnL}
\left( {{\cal P}^2 x^2  + {\cal Q}^2 } \right)^2 \left( {\frac{{dL\left( x \right)}}{{dx}}} \right)^2  + 4{\cal P}^2 {\cal Q}^2 L\left( x \right) + 4M^2 \left( {{\cal P}^2  - {\cal Q}^2 } \right) = 0\,.
\ee

One can find that the vector solution
\be
L\left( x \right) = \frac{2C_1 {\cal PQ}x + C_2 ({\cal P}^2x^2-{\cal Q}^2)}{({\cal Q}^2+{\cal P}^2x^2)}
\ee 
where $C_1$ dan $C_2$ are some constants which are related by 
\be 
{\cal QP}C_2^2  = m^2 \left( {{\cal Q}^2  - {\cal P}^2 } \right) - {\cal P}^2 {\cal Q}^2 C_1^2 \,.
\ee

\section{Direct calculation for central charge}

Following the general formula for the central charge given in \cite{Hartman:2008pb}\footnote{We give an alternative computation by following \cite{Mei:2010wm} in appendix}, where for a general form of near horizon metric
\be 
ds^2  = \Gamma \left( \theta  \right)\left( { - y^2 d\tau^2  + y^{ - 2} dy^2  + \alpha \left( \theta  \right)d\theta ^2 } \right) + \gamma \left( \theta  \right)\left( {d\varphi  + kyd\tau} \right)^2 
\ee
and a general form of Maxwell fields
\be 
A_\mu  dx^\mu   = f\left( \theta  \right)\left( {d\varphi  + kyd\tau} \right)\,,
\ee 
the corresponding central charge given from the spacetime diffeomorphism is
\be \label{c-grav}
c_{grav}  = 3k\int\limits_0^\pi  {\left( {\Gamma \left( \theta  \right)\alpha \left( \theta  \right)\gamma \left( \theta  \right)} \right)^{1/2} } 
\ee 
while the symmetries associated to the gauge field $A_\mu$ do not contribute to the total central charge.

Using the formula (\ref{c-grav}) to the near horizon metric (\ref{metric-nhMKnotglobal}) finally gives us a central charge
\be \label{c-MK-1}
c_{grav}  = 12M^2 \left( {1 - M^4 B^4 } \right)\,.
\ee
In the absence of parameter $B$, the central charge above reduces to the one computed in extreme Kerr black holes \cite{Guica:2008mu}, as expected. 

\section{Ernst potentials for Kerr-Newman}\label{app.Ernst}

It is known that Kerr solution is the electrically neutral limit of the Kerr-Newman solution. Accordingly, in the Ernst formalism in getting the stationary solution with axial symmetry in Einstein-Maxwell theory, the Ernst potential(s) which lead to the Kerr metric can also be obtained from the ones which give Kerr-Newman line element. In Boyer-Lindquist coordinates $({\tilde t},{\tilde r},\theta,{\tilde \phi})$, the Ernst potentials for Kerr-Newman solution read \cite{Ernst}
\be 
{\cal E} =  - \left( {{\tilde r}^2  + a^2  - \frac{{2Ma^2 + ia\left( {2M{\tilde r} - Q^2 } \right)\cos \theta }}{{{\tilde r} + ia\cos \theta }}} \right)\sin ^2 \theta - \frac{{\left( {4Ma + iQ^2 \cos \theta } \right)\left( {a - i{\tilde r}\cos \theta } \right)}}{{\left( {{\tilde r} + ia\cos \theta } \right)}}\,,
\ee
and
\be
\Phi  = \frac{{Q\left( {a - i{\tilde r}\cos \theta } \right)}}{{\left( {{\tilde r} + ia\cos \theta } \right)}}\,.
\ee
Setting $Q = 0$ in the potentials above, which yields $\Phi=0$, gives the Ernst potential for Kerr solution.

\section{Negative temperature}\label{app.NegativeTemp}

When we are discussing the kinetic theory of gases, the phrase ``negative temperature'' is counterintuitive, since we know that the kinetic energy of particle has no bound. However, in spin systems, the magnetic energy of the system is bounded from above \cite{Negative-Temp}, which is the typical of a system where the negative temperature can exist. Such situation is reached when the states with higher energy is more occupied the ones with lower energy. Moreover, from the relation between absolute temperature $T$, entropy $S$, and internal energy $U$,
\be 
\frac{1}{T} = \left( {\frac{{\partial S}}{{\partial U}}} \right)_Q \,,
\ee 
we can learn that when the temperature is negative, the system could still gain more entropy while releasing its internal energy \cite{Tahir-Kheli}. In the last equation, we use the subscript $Q$ to denote that the partial differentiation of entropy function $S$ with respect to the internal energy $U$ is taken when the thermodynamical variable $Q$ is kept to be fixed.

\end{document}